\documentclass[preprint,10pt]{elsarticle}

\usepackage{graphicx}
\usepackage{amssymb}
\usepackage[usenames,dvipsnames]{color}
\usepackage{lineno}
\DeclareMathSizes{10.95}{10}{7}{7} 

\journal{Journal Name}

\begin{document}

\begin{frontmatter}


\title{Some cosmological models coming from gravitational theories having torsional degrees of freedom}



\author[dcf,usp]{J. Lorca Espiro}
\ead{javier.lorca.espiro@ifusp.br}
\author[uls]{Yerko V\'asquez}
\ead{yvasquez@userena.cl}
\address[dcf]{Departamento de Ciencias F\'\i sicas, Facultad de Ingenier\'\i a, Ciencias y
Administraci\'on, Universidad de La Frontera, Avda. Francisco Salazar 01145,
Casilla 54-D Temuco, Chile.}
\address[usp]{Departamento de F\'\i sica Matem\'atica do Instituto de F\'\i sica Universidade de S\~{a}o Paulo
CP 66.318 05314-970, S\~{a}o Paulo, SP Brasil.}
\address[uls]{Departamento de F\'\i sica, Universidad de la Serena, Avenida Cisternas 1200, La Serena, Chile.}

\begin{abstract}
In this work we consider gravitational theories in which the effect of coupling characteristic classes, appropriately introduced as operators in the Einstein-Hilbert action, has been taken into account. As it is well known, this approach strays from the framework of general relativity since it results in theories in which torsion can be present.
We consider here all the characteristic classes that are consistent with a four-dimensional space-time manifold. 
Then, we present explicit expressions for the contortion $1$-form and torsion $2$-form for a broad class of conditions in various cases of interest. Additionally, we use the same framework to study cosmological scenarios that are obtained mainly by selecting the flat FLRW metric and an ideal fluid. 
\end{abstract}


\end{frontmatter}


\section{INTRODUCTION}
\label{Intro}

In recent years, the interest on gravitational theories with torsional degrees of freedom has steadily grown insomuch as they have exhibited interesting cosmological implications, as witnessed in the study of missing matter problems, as well as providing a new mechanism to explain the acceleration of the Universe based on modifications of general relativity (GR), instead of introducing an exotic content of matter.

It is well known that in the Einstein--Cartan theory the source of torsion is the spin current and hence the curvature and torsion become independent degrees of freedom of the gravitational field; however, in this theory, torsion has no dynamics and in four space-time dimensions it vanishes in the absence of sources. To overpass this, one can or add higher order corrections to the curvature or couple additional fields to gravity, resulting in theories with propagating torsion. Among the theories of gravity that encompass torsion, we can mention the Poincar\'{e} gauge theory of gravity, teleparallel gravity and $f(T)$-gravity, being the last two solely based on torsion. The introduction of torsion induces new physical effects and modifies the local degrees of freedom of the theory. For instance, the Einstein-Hilbert theory can be interpreted as a reduction of a higher dimensional model of the Chern-Simons type in which a propagating torsion solution naturally appears \cite{Zumino,Banados}. The addition of torsion also enhances the possibility of an accelerated expansion of the universe among others scenarios \cite{Buchbinder,Poplawski,Capozziello,Boehmer}. An authoritative review to the literature of space-time with torsion in cosmology is provided in \cite{Puetzfeld:2004yg}. Also, it has been proposed that the appearance of torsion would induce observable effects over the neutrino oscillation \cite{Adak}.

The theory considered here has no spin densities, but there is a propagating torsion due to the coupling of characteristic classes with scalar fields. We only consider characteristic classes consistent with a four-dimensional space-time manifold structure, i.e., those constructed from the rational cohomology ring, namely the \textit{Euler} and the \textit{Pontryagin} classes \cite{donaldson}, as well as the Chern type \textit{Nieh-Yan} class \cite{Nieh:1981ww,Nieh:2007zz} which is the only density immediately null in the absence of torsion, see also \cite{Li,Date:2008rb}. 
The effects of adding the aforementioned classes can be acknowledged for by introducing a new term into the connection $1$-form called contortion, which in turn is responsible for the torsion, see for instance \cite{Baekler}.

An effective extension of GR by coupling the Pontryagin invariant with a (non-dynamical) scalar field in absence of torsion was studied in the literature on the so-called  \textit{Chern-Simons gravity} \cite{Jackiw:2003pm}, whereas a generalization of this theory, in which the torsion was included, was considered in \cite{Cantcheff:2008qn,Alexander:2009tp} and the references therein, the theory being mainly motivated by anomaly cancellation in particle physics and string theory. In Ref. \cite{Leigh} four-dimensional gravity deformed by the Nieh-Yan invariant coupled to a scalar field was studied, and an exact solution to the theory was obtained, the so-called \textit{torsion vortex}, which has an interesting holographic interpretation due to the $AdS/CFT$ correspondence. Furthermore, cosmological models of a theory where the Euler invariant was coupled to scalar field have been considered in  \cite{Toloza:2012sn}, see also \cite{Toloza}, the coupling of the scalar field to the Euler invariant can also arise from compactifications of a higher-dimensional theory of the Lovelock family \cite{Mardones:1990qc}. On the other hand, slowly rotating black hole solutions to Chern-Simons gravity in the presence of torsion was presented in \cite{Cambiaso}.

A somewhat similar approach to the one presented here has been used in \cite{Kaul,Sengupta} were the focus was to study the effects of these couplings over the topology and differential structure of the associated symplectic manifold. This was driven by having in mind the later canonical quantization of the theory. Quantizing falls out of the scope of the present work 
and we refer the reader to \cite{Shapiro:2001rz,Kaul:2014jka,Sengupta:2015gva}  and the references therein for quantum aspects of theories with torsion.


This paper aims to consider gravitational theories beyond GR by allowing torsional degrees of freedom coming from all the characteristic classes known to be compatible with a four-dimensional space-time structure. These theories are first studied in a purely gravitational background setting, which results in obtaining analytical expressions for the important dynamical variables in the fashion of the first order formalism for all the cases considered.  We then study several examples in a cosmological scenario, which may be capable to explain different stages of cosmic evolution without the need to introduce an exotic content of matter. The structure of the paper is as follows: in Section \ref{gen_cons_sec} we introduce some mathematical background required in the body of the work. In Section \ref{cas_sec} we obtain explicit expressions for the torsion $2$-form and the contortion $1$-form for four different cases. In Section \ref{cos_sol_sec} we present the appropriate changes that must be carried out for the cosmological studies of these field equations. Consequently, we show several examples in which solutions have been obtained. Finally, we close with some remarks of the solutions shown on the body of the text.

\section{GENERAL CONSIDERATIONS}
\label{gen_cons_sec}

We begin by summarizing the standard vierbein formalism and the basics of differential geometry needed in the next sections, see for instance  \cite{Dona,Giulini,nakahara,nash,hatcher}. On a four-dimensional space-time manifold $\mathcal{M}$ with a cosmological constant $\Lambda$, the Einstein-Hilbert action can be written as
\begin{equation}\label{EHaction}
{S_{EH}}[e^c, \omega^a_{\;\;b}] = \frac{1}{\kappa }\int\limits_{\mathcal{M}}{ {{\epsilon_{abcd}}{R^{ab}} \wedge{e^c} \wedge {e^d} + \frac{\Lambda }{6}{\epsilon _{abcd}}{e^a} \wedge {e^b} \wedge {e^c} \wedge {e^d}} } ,
\end{equation}
where $\kappa=32 \pi G$ is the Newton constant, $e^a$ denotes the $1$-form frame fields or vierbein  and $\omega^a\,_b$ is a general connection $1$-form, the latter defines the curvature to be 
\begin{equation}\label{originalcurvature}
R^{ab} = d \omega^{ab} + \omega^a\,_c \wedge \omega^{cb}.
\end{equation}

The Palatini variation of the action (\ref{EHaction}) gives the Einstein's field equations and also, as an independent constraint, the null torsion condition
\begin{equation}\label{originaltorsion}
T^a = de^a+\omega^a\,_b \wedge e^b=0,
\end{equation}
allowing to write the action as well as the connection entirely in terms of the vierbein (or ultimately in terms of the metric) as
\begin{equation}\label{originalconnection}
\omega^{ab} = \frac{1}{2} \left[ i^b \left(d e^a\right) - i^a \left( d e^b \right) + i^{ab} \left( d e_c\right)e^c \right],
\end{equation}
where $d\left( \cdot \right)$ is the exterior derivative and $i^c \left( \cdot \right) \equiv i^{e_c} \left( \cdot \right) $ is the interior or slant product with respect to the vierbein. We have also denoted $i^{ab} \left( \cdot \right) = -i^{ba} \left( \cdot \right) =i^a \left( i^b \left( \cdot \right) \right)$. Eq. (\ref{originalconnection}) defines the connection $1$-form for GR, also known as the Levi-Civita connection. We can think of $\mathcal{M}$ as being a $SO\left( 4 \right)$-bundle or a $SO\left(3,1\right)$-bundle space-time, depending on the signature of the metric.

For an oriented real four-manifold $\mathcal{M}$ the characteristic classes available comprise the Stiefel-Whitney classes $w_i \left( T \mathcal{M} \right) \in H^i \left( \mathcal{M} ; \mathbb{Z}/2 \right)$ and the Euler and Pontryagin classes $e \left( \mathcal{M} \right), \, p_1 \left( T \mathcal{M} \right) \in H^4 \left( \mathcal{M} ;\mathbb{Z}\right) = \mathbb{Z}$ \cite{donaldson}. By allowing the four-manifold $\mathcal{M}$ to be an almost complex smooth four-manifold the Chern classes $c_i \left( \mathcal{M} \right) = c_i \left( T \mathcal{M} \right) \; \in H^{2i} \left( \mathcal{M} ,\mathbb{Z} \right)$ can also be considered \cite{donaldson}. Among the previous possibilities we will specifically use the \textit{Pontryagin} and \textit{Euler} densities, as well as the Chern type  \textit{Nieh-Yan} density which has been proven to be important to study the spin structure of the theory in a quantum context \cite{Kaul,Sengupta,Mercuri:2010yj,Sengupta:2009kg,Liko}. These have, locally, the following representations as $4$-forms, respectively.
\begin{eqnarray}
\nonumber C_{P} & = & d \, \left( \, {\tilde C}_{P} \, \right) \\
\label{Pontryagin}  & = &  d \left( \omega^a_{\;\;b} \wedge \left[ R^b_{\;\;a} - \frac{1}{3} \omega^b_{\;\;c} \wedge \omega^c_{\;\;a} \right] \right) = R^a_{\;\;b} \wedge R^b_{\;\;a} ,\\
\nonumber C_{E} & = & d \, \left( \, {\tilde C_{E}} \, \right) \\
\label{Euler} & = & \epsilon_{abcd} d \left( \omega^{ab} \wedge \left[ R^{cd} - \frac{1}{3} \omega^c_{\;\;f} \wedge \omega^{fc}\right] \right) = \epsilon_{abcd} \; R^{ab} \wedge R^{cd},\\
\nonumber C_{NY} & = & d \, \left( \, {\tilde C}_{NY} \, \right) \\
\label{NiehYan} & = &  d \left( T^a \wedge e_a \right) = T^a \wedge T_a - R_{ab} \wedge e^a \wedge e^b ,
\end{eqnarray}
where we have denoted by a tilde the corresponding $3$-form associated with each exact $4$-form presented above in an obvious way.

From now on, we will denote the connection of GR (Levi-Civita connection) as $\bar{\omega}^{ab}$, so that a generalized spin connection can split in the following form \cite{Leigh,Cambiaso,Cho,Hatzinikitas,Petkou,Chandia,Randal}
\begin{equation}
\label{connection} \omega^{ab} = \bar \omega^{ab}+K^{ab},
\end{equation}
where $K^{ab}$ is called the \textit{contortion}. Therefore, the torsion 2-form can be written as 
\begin{equation}
\label{contortion} T^a  =  d_{\omega} e^a =K^{ab} \wedge e_b,
\end{equation}
where $d_{\omega}$ is the covariant derivative with respect to the connection $\omega^{ab}$.

Hence, the contortion $K^{ab}$ is the object responsible for the presence of torsion $T^a$ as anticipated. Similarly, using (\ref{originalcurvature}) and (\ref{connection}) the curvature $2$-form can be written in the following way
\begin{equation}
\label{Riemanntensorgral}  R^{ab} \, = \, \bar{R}^{ab} + d_{\omega} K^{ab} - K^a_{\;\;c} \wedge K^{cb},
\end{equation}
where $\bar{R}^{ab} \doteq d \bar{\omega}^{ab} + \bar{\omega}^a_{\;\;c} \wedge \bar{\omega}^{cb}$. Finally, the following, known as the Bianchi identities
\begin{eqnarray}
\label{Bianchi1} d_{\omega} T^a & = & R^a_{\;\;b} \wedge e^b,\\
\label{Bianchi2} d_{\omega} R^a_{\;\;b} & = & 0,
\end{eqnarray}
are satisfied.

\subsection{The full action}
\label{act_sec}

It should be clear, following the discussion in the previous section, that knowing the Levi-Civita connection $\bar \omega_{ab}$ and the contortion $K_{ab}$ $1$-form is sufficient to describe all the physical observables of the theory. 
Hence, the problem narrows down to find an expression for $K_{ab}$. 
Let us address this by considering the following family of actions
\begin{equation}
\label{quasiaction} S \left[ {{e^c},\omega^a_{\;\;b}}, \alpha_i \right] = {{S_{EH}}\left[ {{e^c},\omega^a_{\;\;b}} \right]} - \frac{1}{\kappa} \int\limits_\mathcal{M} \left( \tau {C_{P}} + \varphi {C_{NY}}  + \chi {C_E}\right)  ,
\end{equation}
where we have introduced the dimensionless fields $\tau,\, \chi$, and the field $\varphi$ with dimensions $[ l^{-2} ]$ (collectively written as $\alpha_i$) that serve as couples to the characteristic classes (\ref{Pontryagin}), (\ref{Euler}) and (\ref{NiehYan}) respectively.
The coupling $\tau$ and $\varphi$ are thought as being axionic fields whereas $\chi$ is a scalar field introduced in the family of actions (\ref{quasiaction}) to generate torsion.

Note also that we can write the action (\ref{quasiaction}) as
\begin{eqnarray}
\nonumber S \left[ {{e^c},\omega^a_{\;\;b}}, \alpha_i \right] & = & {{S_{EH}}\left[ {{e^c},\omega^a_{\;\;b}}, \alpha_i \right]} + 
\frac{1}{\kappa} \int\limits_\mathcal{M} {\left( {d\tau  \wedge {{\tilde C}_{P}} + d\varphi  \wedge {{\tilde C}_{NY}} + d\chi  \wedge {{\tilde C}_E}} \right)} \\
\label{action1}  & & - \frac{1}{\kappa} \int\limits_{\partial \mathcal{M}} \left( \tau {\tilde{C}_{P}} + \varphi \, {\tilde{C}_{NY}} + \chi \,{\tilde{C}_E} \right),
\end{eqnarray}
where we used Stokes theorem to obtain the last term. Note that the action has several non-trivial contributions on the boundary that consequently give rise to several possibilities we need to analyze if we want to avoid extra complications when extremizing the action,

\begin{itemize}



\item $\alpha_i$ is a parameter 

Since the last term is an exact form it will not have contributions over the field equations. However, for a compact manifold with boundary there are non-linear higher derivative contributions coming from the characteristic densities acting over the boundary. If this is considered leads to nontrivial results in the context of AdS/CFT correspondence which are beyond the scope of the present work. The latter was studied in the following references \cite{Aros,Miskovic}. Also, in \cite{Aros:1999kt} it was shown that the Euler index appears in the Lagrangian for asymptotically locally AdS (ALADS) manifolds and in \cite{Sengupta:2013lxa} it was shown the emergence of all three topological invariants for ALADS manifolds.

\item $\alpha_i$ is a map

Again, since the last term is an exact differential it will not have contributions on the field equations. However, this term will have nontrivial boundary conditions depending on the topology of the manifold considered. For instance, if $\mathcal{M}$ is a compact manifold the $\alpha_i$ maps are bounded with Neumann type of boundary conditions \cite{Lambert}. Since considering a boundary implies to adjust the boundary terms to behave like a generalized Gibbons-Hawking-York type of term \cite{Dyer}, we will work with the particular case of closed manifolds (compact and without boundary) to avoid unnecessary discussion on the matter. Furthermore, this is consistent with the FLRW metric chosen to study cosmological scenarios in the subsequent sections.

\end{itemize}


\section{EXPLICIT EXPRESSIONS FOR THE TORSION AND CONTORTION IN DIFFERENT TOPOLOGICAL SETUPS}
\label{cas_sec}

Before tackling the problem described by the action introduced in section \ref{act_sec} (Eq. (\ref{action1})), we propose few simplified cases in which we analyze the separated effect of the topological terms. We first take a look at the two simplest scenarios, where we consider the contributions from the Nieh-Yan and Pontryagin terms, and Pontryagin and Euler terms respectively; for these two cases in fact we can obtain explicit expressions. Consequently, we present two more cases that we approach in a more heuristic way. The latter cases describe the contribution of the Nieh-Yan and Euler terms and the last case in which all three terms are allowed to be present. 
As previously stated, we start by varying the action  (\ref{action1}) considering $\alpha_i$ to be maps, i.e. couplings as $0$-forms and $\mathcal{M}$ a compact manifold.

\subsection{Case 1: Coupling Nieh-Yan and Pontryagin densities to the Einstein-Hilbert action}

This first case can be carried out very straightforwardly. The field equations are
\begin{eqnarray}
\delta e & : & \begin{array}{ccl}
 0 & = & - \epsilon_{abcd}e^b \wedge R^{cd} - \frac{\Lambda}{3} \epsilon_{abcd}e^b \wedge e^c \wedge e^d + d \varphi \wedge T_a,\
\end{array} \label{fe11} \\
\delta \omega  & : & \begin{array}{ccl}
0 & = & - 2 \epsilon^{abcd} T_c \wedge e_d -2d\tau \wedge R^{ab} -d\varphi \wedge e^a \wedge e^b,\
\end{array} \label{fe12} \\
\delta \varphi & : & \begin{array}{cccl}
0 & = & T^a \wedge T_a - R_{ab} \wedge e^a \wedge e^b,\
\end{array} \label{fe13} \\
\delta \tau & : & \begin{array}{cccl}
0 & = & R^a_{\;\;b} \wedge R^b_{\;\;a}.
\end{array} \label{fe14}
\end{eqnarray}
In order to solve this set of equations, we need to exclude the curvature and find an explicit expression for the contortion.
This can be done by multiplying $d\tau \wedge ( \cdot ) $ over Eq. (\ref{fe11}). After, Eq. (\ref{fe12}) can be used to obtain
\begin{equation}
2 \sigma e^b \wedge T_b\wedge e_a + d\tau \wedge d\varphi \wedge T_a -\frac{1}{2} \epsilon_{abcd} e^b \wedge e^c \wedge e^d \wedge (d\varphi-\frac{2 \Lambda}{3} d\tau)=0,
\end{equation}
where $\sigma=1$ for an Euclidean metric with signature $(+,+,+,+)$ and $\sigma=-1$ for a Lorentzian metric with signature $(+,-,-,-)$.
For simplicity, we will consider the condition $d\tau \wedge d\varphi=0$, thus, the above equation reduces to
\begin{equation}
2 \sigma e^b \wedge T_b\wedge e_a-\frac{1}{2} \epsilon_{abcd} e^b \wedge e^c \wedge e^d \wedge (d\varphi-\frac{2 \Lambda}{3} d\tau)=0,
\end{equation}
and applying the interior product $i^a \left( \cdot \right)$ to this equation, we get
\begin{equation}
e^b\wedge \left(-4 \sigma T_b+ \epsilon_{abcd} e^c \wedge e^d i^a (d \varphi-\frac{2 \Lambda}{3} d\tau)\right)=0,
\end{equation}
thus, the torsion is obtained straightforwardly to yield
\begin{equation}\label{torsion1}
T_a^{\left( 1 \right)} = -\frac{1}{4 \sigma}{\epsilon _{abcd}}{e^b} \wedge {e^c}\left( {{\mathcal{L}^d}\left( \varphi  \right)-\frac{2\Lambda}{3}\mathcal{L}^d}\left( \tau  \right) \right) +  \delta d \theta_1 \wedge e_a,
\end{equation} 
where $\delta$, $\theta_1: \mathcal{M} \rightarrow \mathbb{R}$ are $0$-forms that need to be adjusted in order to fulfill the rest of the restrictions imposed by the field equations.
On the other hand, multiplying Eq. (\ref{fe11}) by $d \varphi $ leads to
\begin{equation}\label{curvature1ansatz}
\epsilon _{abcd}\left( R^{cd}+\frac{\Lambda }{3}e^{c}\wedge e^{d}\right) \wedge e^{b}\wedge d\varphi = 0,
\end{equation}
which in turn suggests the following expression for the curvature $2$-form
\begin{equation} \label{curvgamma0}
\nonumber R^{\left( 1 \right)}_{ab}  =  -\frac{\Lambda}{3} e_{a}\wedge e_{b} + d\tau \wedge A_{ab}+d\varphi \wedge B_{ab} + \gamma \epsilon _{abcd}e^{c} \wedge e^{d},
\end{equation}
where $\gamma$ is an arbitrary function and $A_{ab}$, $B_{ab}$ are 1-forms that must be determined from the remaining field equations. Additionally, by inserting Eqs. (\ref{torsion1}) and (\ref{curvgamma0}) in (\ref{fe12}) yields $\delta d\theta_1=-\gamma d\tau$. 
The contortion is obtained using (\ref{torsion1}) and (\ref{contortion})
\begin{equation}\label{contortionfinal1}
K^{\left( 1 \right)}_{ab} = \frac{1}{4 \sigma} {\epsilon_{abcd}}{e^c}\left({\mathcal{L}^d} \left( \varphi \right)-\frac{2\Lambda}{3} {\mathcal{L}^d} \left( \tau \right)  \right) - \gamma \{ \mathcal{L}_a \left( \tau \right) e_b - \mathcal{L}_b \left( \tau \right) e_a \},
\end{equation}
which is manifestly skew-symmetric in the indices $a$ and $b$. Note that by equations (\ref{torsion1}) and (\ref{contortionfinal1}) we immediately recognize that $\left(\delta d \theta_1 + \gamma d \tau \right) \wedge e_a = 0$. On the other hand, inserting Eq. (\ref{curvgamma0}) in Eq. (\ref{fe11}), we find that the following expressions satisfy the resulting equation
\begin{eqnarray}
\nonumber A_{ab} & = & \alpha \epsilon_{abcd}\mathcal{L}^c(\tau)e^d + \beta \epsilon_{abcd} \mathcal{L}^c(\varphi)e^d, \\
B_{ab} & = & \nonumber B_{ab}^{(1)}+B_{ab}^{(2)}, \\
\nonumber B_{ab}^{(1)} & = & \hat{\alpha} \epsilon_{abcd}\mathcal{L}^c(\tau)e^d + \hat{\beta} \epsilon_{abcd} \mathcal{L}^c(\varphi)e^d, \\
\label{cond1} B_{ab}^{(2)} & = & \frac{1}{8 \sigma} \left\{ e_{a} i_{b} \left(-\frac{2\Lambda }{3} d\tau + d\varphi\right)  -e_{b} i_{a} \left(-\frac{2\Lambda }{3} d\tau + d\varphi \right) \right\},
\end{eqnarray}
where we have inserted $\alpha, \beta, \hat{\alpha}, \hat{\beta},\delta:\mathcal{M} \rightarrow \mathbb{R}$, functions or $0$-forms that must depend on the rest of the parameters and couplings only. However, notice that the expressions of the 1-forms $A_{ab}$ and $B_{ab}$ were not actually necessary for find the explicit expression for the torsion. Notice also, that $B_{ab}$, in contrast to $A_{ab}$, acquires an additional term due to the last term of Eq. (\ref{fe11}). Then, taking into considerations restrictions (\ref{fe13}) and (\ref{fe14}) we obtain
\begin{eqnarray}
\nonumber 0 & = & \frac{1}{2\sigma} \delta d\theta_1\wedge \epsilon_{abcd}e^a \wedge e^b \wedge e^c \wedge \mathcal{L}^d \left( \varphi-\frac{2\Lambda}{3} \tau \right)+ d\tau \wedge A_{ab} \wedge e^a \wedge e^b + \\
\nonumber && + d\varphi \wedge B_{ab}^{(1)} \wedge e^a \wedge e^b + \gamma \epsilon_{abcd} e^a \wedge e^b \wedge e^c \wedge e^d, \,\,\,\,\,\,\, \\
 0 & = & \gamma  d\varphi \wedge \epsilon_{abcd} e^a \wedge e^b \wedge B^{(2) cd}.
\end{eqnarray}
The restrictions are not very illustrative for our purposes mainly because we need to use a definite metric and therefore its analysis will be postponed until section \ref{cos_sol_sec}.

\subsection{Case 2: Coupling Nieh-Yan and Euler characteristic invariants to the Einstein-Hilbert action}

The second case we consider is also quite straightforward. By imposing $\tau = 0$ in the action (\ref{action1}) the following field equations are obtained
\begin{eqnarray}
\delta e & : & \begin{array}{ccl}
0 & = & -\epsilon _{abcd}e^{b}\wedge R^{cd}-\frac{\Lambda}{3}\epsilon_{abcd}e^{b}\wedge e^{c}\wedge e^{d}+d\varphi \wedge T_{a},\
\end{array} \label{fe21} \\
\delta \omega &:& \begin{array}{ccl}
0 & = & -2\epsilon_{abcd}T^{c} \wedge e^{d}-d \varphi \wedge e_{a} \wedge e_{b} - 2\epsilon _{abcd} d \chi \wedge R^{cd},\
\end{array} \label{fe22}\\
\delta \chi & : & \begin{array}{ccl}
0 & = &   \epsilon_{abcd} R^{ab} \wedge R^{cd},
\end{array}\label{fe23}\\
\delta \varphi & : & \begin{array}{cccl}
0 & = & T^a \wedge T_a - R_{ab} \wedge e^a \wedge e^b,\
\end{array} \label{fe24}
\end{eqnarray}
with the last equation coinciding with Eq. (\ref{fe13}), as expected. Here, the approach goes along the lines of what is presented in the previous subsection. Here we multiply $d\chi \wedge \left( \cdot \right) $ to Eq. (\ref{fe21}). After combining this result with Eq. (\ref{fe22}), we get
\begin{equation}\label{fe25}
-\frac{\Lambda}{3}{\epsilon_{abcd}}d\chi  \wedge {e^b} \wedge {e^c} \wedge {e^d} + d \chi  \wedge d\varphi  \wedge {T_{a}} - {\epsilon_{abcd}} {e^b}\wedge {T^c} \wedge {e^d} = 0.
\end{equation}
In order to find an explicit expression for the torsion, we will consider $d\chi \wedge d\varphi=0$, so, the above equation simplifies to
\begin{equation}
{\epsilon_{abcd}}{e^b} \left( {T^c}-\frac{\Lambda}{3}d\chi  \wedge {e^c} \right) \wedge {e^d}= 0,
\end{equation}
and the torsion is obtained straightforwardly to yield
\begin{equation}\label{torsion2}
T_a^{(2)}  =  {\frac{\Lambda }{3} }d\chi  \wedge {e_a}+\delta \epsilon_{abcd} e^b \wedge e^c \mathcal{L}^d \left( \theta_2 \right),
\end{equation}
where $\delta$, $\theta_2: \mathcal{M} \rightarrow \mathbb{R}$ are $0$-forms which must be adjusted using the field equations. As in the previous section, when multiplying Eq. (\ref{fe21}) by $d \varphi $ we obtain
\begin{equation}\label{curvatuansatz}
\epsilon _{abcd}\left( R^{cd}+\frac{\Lambda }{3}e^{c}\wedge e^{d}\right) \wedge e^{b}\wedge d\varphi = 0,
\end{equation}
which, in turn, suggests the following expression for the curvature $2$-form
\begin{equation} \label{curvgamma1}
\nonumber R^{\left( 2 \right)}_{ab}  =  -\frac{\Lambda}{3} e_{a}\wedge e_{b} + d\chi \wedge A_{ab}+d\varphi \wedge B_{ab} + \gamma \epsilon _{abcd}e^{c} \wedge e^{d},
\end{equation}
where $\gamma$ is an arbitrary function and $A_{ab}$, $B_{ab}$ are 1-forms that must be determined from the remaining field equations. Additionally, inserting Eqs. (\ref{torsion2}) and (\ref{curvgamma1}) in Eq. (\ref{fe22}) yields

\begin{equation}
\delta d\theta_2=-\frac{1}{4 \sigma} \left( d\varphi+ 8 \sigma \gamma d\chi \right).
\end{equation}
Finally, with the aid of Eq. (\ref{contortion}) and Eq. (\ref{torsion2}) we find the contortion to be
\begin{equation}\label{contortionfinal2}
{{K}^{(2)}_{ab}} = {\frac{\Lambda }{3} } \left\{ {\mathcal{L}_a} \left(\chi \right){e_b} - \mathcal{L}_b \left( \chi \right){e_a} \right\} +\frac{1}{4 \sigma} \epsilon_{abcd} e^c \left( \mathcal{L}^d \left( \varphi \right)+8 \sigma \gamma  \mathcal{L}^d \left( \chi \right)  \right),
\end{equation}
which is manifestly skew-symmetric in the indices $a$ and $b$. On the other hand, inserting Eq. (\ref{curvgamma1}) in Eq. (\ref{fe21}), we find that the following expressions satisfy the resulting equation
\begin{eqnarray}
\nonumber A_{ab} & = & \alpha \epsilon_{abcd}\mathcal{L}^c(\chi)e^d + \beta \epsilon_{abcd} \mathcal{L}^c(\varphi)e^d, \\
 \nonumber B_{ab} & = & B_{ab}^{(1)}+B_{ab}^{(2)}, \\
\nonumber B_{ab}^{(1)} & = & \hat{\alpha} \epsilon_{abcd}\mathcal{L}^c(\chi)e^d + \hat{\beta} \epsilon_{abcd} \mathcal{L}^c(\varphi)e^d, \\
\label{cond2} B_{ab}^{(2)} & = & \frac{1}{8 \sigma} \left\{ e_{a} i_{b} \left(d\varphi +8 \sigma \gamma d\chi \right)  -e_{b} i_{a} \left(d\varphi + 8 \sigma \gamma d\chi \right) \right\},
\end{eqnarray}
where $\alpha, \beta, \hat{\alpha}, \hat{\beta},\delta:\mathcal{M} \rightarrow \mathbb{R}$ are arbitrary functions. Now, when taking into consideration the restrictions coming from Eqs. (\ref{fe23}) and (\ref{fe24}) we obtain
\begin{eqnarray}
\nonumber 0 & = & \left( \frac{\Lambda^2}{9} + 4 \sigma \gamma^2 \right) \epsilon_{abcd} e^a \wedge e^b \wedge e^c \wedge e^d + 8 \sigma \gamma (d\chi \wedge A_{ab} \wedge e^a \wedge e^b \\
\nonumber && +d\varphi \wedge B_{ab}^{(1)} \wedge e^a \wedge e^b)-\frac{2 \Lambda}{3} d\varphi \wedge \epsilon_{abcd} e^a \wedge e^b \wedge B^{(2) cd}, \\
\nonumber 0 & = & -\frac{2\Lambda}{3} \delta d\chi \wedge \epsilon_{abcd} e^a \wedge e^b \wedge e^c \wedge \mathcal{L}^d(\theta_2)+ d\chi \wedge A_{ab} \wedge e^a \wedge e^b+\\
&&d\varphi \wedge B_{ab}^{(1)} \wedge e^a \wedge e^b+ \gamma \epsilon_{abcd} e^a \wedge e^b \wedge e^c \wedge e^d, \,\,\,\,\,\,\,
\end{eqnarray}
respectively. As before, these restrictions are not very illustrative for our purposes mainly because we need to use a definite metric and therefore its analysis will be postponed until the section \ref{cos_sol_sec}. However, notice that for $\Lambda=0$ the above constraints impose $\gamma=0$.

\subsection{Case 3: Pontryagin and Euler characteristic invariants added to the Einstein-Hilbert action}

Settting $\varphi=0$ in (\ref{action1}) yields the following field equations
\begin{eqnarray}
\label{fe31} \delta e & : & \begin{array}{ccl}  
0 & = & -\epsilon _{abcd}e^{b}\wedge R^{cd}-\frac{\Lambda }{3}\epsilon_{abcd}e^{b}\wedge e^{c}\wedge e^{d} ,\
\end{array} \\
\label{fe32} \delta \omega & : & \begin{array}{ccl}
0 & = & -\epsilon_{abcd}T^{c}\wedge e^{d} - d\tau \wedge R_{ab} -\epsilon _{abcd} d \chi \wedge R^{cd},\
\end{array}\\
\label{fe33} \delta \tau & : & \begin{array}{cccl}
0 & = & R^a_{\;\;b} \wedge R^b_{\;\;a}\ ,
\end{array} \\
\label{fe34} \delta \chi & : & \begin{array}{ccl}
0 & = &  \epsilon_{abcd} R^{ab} \wedge R^{cd}.
\end{array}
\end{eqnarray}
Note the presence of the field equations (\ref{fe33}) and (\ref{fe34}) which we have already encountered in the previous cases. The standard approach we used before cannot be applied here directly. In fact, combining Eqs. (\ref{fe31}) and (\ref{fe32}) will lead to expressions that cannot be cast in terms of the contortion and its interior products only. Hence, we take a more heuristic approach: taking the wedge product of $d\chi$ and Eq. (\ref{fe31}) and using (\ref{fe32}) leads to
\begin{equation}
-\epsilon_{abcd}e^b \wedge T^c \wedge e^d-e^b \wedge d\tau \wedge R_{ab}-\frac{\Lambda}{3} \epsilon_{abcd} d\chi \wedge e^b \wedge e^c \wedge e^d=0,
\end{equation}
note that the curvature appears explicitly in the above equation. However, in the following, as we have been doing so far, we will consider $d \tau \wedge d \chi=0 $, thus, Eq. (\ref{fe31}) by itself suggests a curvature of the form
\begin{equation} \label{curvature3ansatz}
\nonumber {R^{\left( 3 \right)}_{ab}} = -\frac{\Lambda }{3}{e_a} \wedge {e_b} + d \tau \wedge A_{ab} + d \chi \wedge B_{ab} + \gamma {\epsilon _{abcd}}{e^c} \wedge {e^d} ,
\end{equation}
where 
\begin{eqnarray}
\nonumber A_{ab} & = & \alpha \epsilon_{abcd}\mathcal{L}^c(\tau)e^d + \beta \epsilon_{abcd} \mathcal{L}^c(\chi)e^d, \\
\label{cond3} B_{ab} & = & \hat{\alpha} \epsilon_{abcd}\mathcal{L}^c(\tau)e^d + \hat{\beta} \epsilon_{abcd} \mathcal{L}^c(\chi)e^d,
\end{eqnarray}
and where $\alpha$, $\beta$, $\hat{\alpha}$ and $\hat{\beta}$ are $0$-forms that must depend on the rest of the parameters and couplings. Note that the exact expressions of the 1-forms $A_{ab}$ and $B_{ab}$ are not actually necessary to find an explicit expression for the torsion, as we will see. Now, by inserting Eq. (\ref{curvature3ansatz}) in Eq. (\ref{fe32}) one obtains
\begin{equation}
\epsilon_{abcd}e^b \wedge \left( T^c+\gamma d \tau \wedge e^c -\frac{\Lambda}{3} d\chi \wedge e^c \right) \wedge e^d=0,
\end{equation}
and from the last expression the torsion is found to be
\begin{equation}\label{cuasitorsion}
{T^{\left( 3 \right)}_a} =  - \left( {\gamma d\tau  - \frac{\Lambda}{3}d\chi } \right) \wedge {e_a} +\delta {\epsilon _{abcd}}{e^b} \wedge {e^c} \wedge {\mathcal{L}^d}\left(\theta_{3} \right) ,
\end{equation}
where $\delta$, $\theta_3: \mathcal{M} \rightarrow \mathbb{R}$ are arbitrary $0$-form. These functions can be adjusted by inserting Eqs. (\ref{cuasitorsion}) and (\ref{curvature3ansatz}) in (\ref{fe32}), yielding
\begin{equation}\label{torsion3previa}
{T^{\left( 3 \right)}_a} =  - \left( {\gamma d\tau  - \frac{\Lambda}{3}d\chi } \right) \wedge {e_a} - \frac{1}{4 \, \sigma}{\epsilon _{abcd}}{e^b} \wedge {e^c} \wedge {i^d}\left({8 \sigma \gamma d\chi  - \frac{{2\Lambda }}{3}d\tau } \right) ,
\end{equation}
where we have denoted $\sigma =1$ for an Euclidean like metric and $\sigma =-1$ for a Lorentzian like metric, as before. The contortion is found to be
\begin{equation}\label{contoprevia}
{K^{\left( 3 \right)}}_{ab} = \gamma (e_a \mathcal{L}_b (\tau)-e_b \mathcal{L}_a (\tau))-\frac{\Lambda}{3} (e_a \mathcal{L}_b (\chi)-e_b \mathcal{L}_a (\chi)) + \frac{1}{4 \, \sigma}{\epsilon_{abcd}}\wedge {e^c} \left({8 \sigma \gamma \mathcal{L}^d(\chi)  - \frac{{2\Lambda }}{3} \mathcal{L}^d(\tau) } \right).
\end{equation}
On the other hand, inserting (\ref{curvature3ansatz}) into Eqs. (\ref{fe33}) and (\ref{fe34}) yields
\begin{eqnarray}
\label{res31} 0 & = & -\frac{2 \Lambda}{3} \left( d\tau \wedge A^{ab} \wedge e_{a} \wedge e_{b}+ d\chi \wedge B^{ab} \wedge e_{a} \wedge e_{b} +\gamma \epsilon _{abcd} e^{a} \wedge e^{b} \wedge e^{c} \wedge e^{d} \right),  \\
\label{res32} 0 & = & \left(\frac{\Lambda^2}{9} +4 \sigma \gamma^2\right) \epsilon_{abcd} e^{a} \wedge e^{b} \wedge e^{c} \wedge e^{d}+8 \sigma \gamma \left( d \tau \wedge A^{ab} \wedge e_{a}\wedge e_{b} +d \chi \wedge B^{ab} \wedge e_{a} \wedge e_{b}\right),\,\,\,\,\,\,\,\,\,\,\,\, 
\end{eqnarray}
where, as before, we have assumed a condition of the type $d \chi \wedge d\tau = 0$ to simplify the aforementioned expressions. Starting from Eq. (\ref{res31}), when combined with Eq. (\ref{res32}), it is easy to obtain 
\begin{equation}\label{res3a}
\gamma =  \pm \frac{ \Lambda }{6 \sqrt{\sigma}},
\end{equation}
for the auxiliary $0$-form, which is valid only for Euclidean solutions, this is $\sigma=1$. 

Inserting this expression back into Eq. (\ref{torsion3previa}) gives
\begin{equation} \label{torsion32}
{T^{\left( 3\right)}_a} =  \frac{\Lambda }{3} \left\{ \left( d \chi {\mp \frac{1}{2} d\tau } \right) \wedge {e_a} + \frac{1}{2}{\epsilon _{abcd}}{e^b} \wedge {e^c} \wedge {\mathcal{L}^d}\left({\tau \mp 2 \chi} \right) \right\},
\end{equation}
which in turn, gives us the corresponding contortion
\begin{equation}
\label{contortionfinal32} {K^{\left( 3 \right)}_{ab}}  = \frac{\Lambda }{3} \left\{ {{\mathcal{L}_a}\left( \chi {\mp \frac{1}{2} \tau }  \right){e_b} - {\mathcal{L}_b}\left(\chi {\mp \frac{1}{2} \tau }  \right){e_a}} \right\} -\frac{\Lambda}{6} {\epsilon _{abcd}}{e^c}{\mathcal{L}^d}\left( {\tau \mp 2\chi} \right).
\end{equation}

Eq. (\ref{contortionfinal32}) have been rearranged in order to highlight the similarities with (\ref{contortionfinal1}) and (\ref{contortionfinal2}). We can see that the heuristic approach paid off. Note that these solutions strongly depend on the fact that the cosmological constant be different from zero.

\subsection{Case 4: Nieh-Yan, Pontryagin and Euler characteristic invariants added to the Einstein-Hilbert action}

The more heuristic approach used on the previous case can also be applied here with some minor adjustments. Let us now consider the entire action (\ref{action1}), and its variation
\begin{eqnarray}
\label{fe41} \delta e & : & \begin{array}{ccl}
0 & = & -\epsilon _{abcd}e^{b}\wedge R^{cd}-\frac{\Lambda }{3}\epsilon_{abcd}e^{b}\wedge e^{c}\wedge e^{d}+d\varphi \wedge T_{a}, \
\end{array}\\
\nonumber \delta \omega & : & \begin{array}{ccl}
0 & = & -2\epsilon_{abcd}T^{c}\wedge e^{d}-2d\tau \wedge R_{ab} - d\varphi \wedge e_{a}\wedge e_{b}-2\epsilon _{abcd} d \chi \wedge R^{cd},
\end{array}\\
\label{fe42}\\
\delta \varphi & : & \begin{array}{cccl}
0 & = & T^a \wedge T_a - R_{ab} \wedge e^a \wedge e^b,\
\end{array} \label{fe43} \\
\delta \chi & : & \begin{array}{ccl}
0 & = &   \epsilon_{abcd} R^{ab} \wedge R^{cd},
\end{array}\label{fe44}\\
\delta \tau & : & \begin{array}{cccl}
0 & = & R^a_{\;\;b} \wedge R^b_{\;\;a}\ .
\end{array} \label{fe45}
\end{eqnarray}
As we have been assuming, for the sake of simplicity, we consider the couplings $0$-forms to satisfy $d\varphi \wedge d\tau = d \varphi \wedge d \chi = d\tau \wedge d \chi=0$. When multiplying Eq. (\ref{fe41}) by $d \varphi $ leads to
\begin{equation}\label{curvature4ansatz}
\epsilon _{abcd}\left( R^{cd}+\frac{\Lambda }{3}e^{c}\wedge e^{d}\right) \wedge e^{b}\wedge d\varphi = 0,
\end{equation}
which in turn suggests the following expression for the curvature $2$-form
\begin{equation} \label{curvgamma}
\nonumber R^{\left( 4 \right)}_{ab}  =  -\frac{\Lambda}{3} e_{a}\wedge e_{b} + d\tau \wedge A_{ab}+d\chi \wedge B_{ab}+ d\varphi \wedge C_{ab} + \gamma \epsilon _{abcd}e^{c} \wedge e^{d},
\end{equation}
where $\gamma$ is an arbitrary function and $A_{ab}$, $B_{ab}$ and $C_{ab}$ are 1-forms that must be determined from the remaining field equations. However, as in the previous cases, it is possible to obtain an explicit expression for the torsion without knowing them. Analogously as in the previous cases, by multiplying Eq. (\ref{fe41}) by $d\chi \wedge \left( \cdot \right)$ combined with Eqs. (\ref{fe42}) and (\ref{curvgamma}), then, inserting this back in (\ref{fe42}) yields the torsion
\begin{equation}\label{torsiopr}
T^{\left( 4 \right)}_{a} = - \left( \gamma d\tau - \frac{\Lambda }{3}d\chi \right) \wedge e_{a} - \frac{1}{4\sigma }\epsilon _{abcd}e^{b} \wedge e^{c}\wedge i^{d}\left( -\frac{2\Lambda }{3} d\tau + d\varphi +8\sigma \gamma d\chi \right),
\end{equation}
and the contortion reads
\begin{eqnarray}
\nonumber K^{\left( 4 \right)}_{ab} & = & \gamma (e_a \mathcal{L}_b (\tau)-e_b \mathcal{L}_a (\tau))-\frac{\Lambda}{3} (e_a \mathcal{L}_b (\chi)-e_b \mathcal{L}_a (\chi)) \\ 
\label{conto} & & + \frac{1}{4 \, \sigma}{\epsilon_{abcd}}\wedge {e^c} \left(- \frac{{2\Lambda }}{3} \mathcal{L}^d(\tau)+ \mathcal{L}^d(\varphi) + {8 \sigma \gamma \mathcal{L}^d(\chi)}   \right).
\end{eqnarray}
Inserting Eq. (\ref{curvgamma}) in Eq. (\ref{fe41}), and based on the results of the previous sections, we get

\begin{eqnarray}
\nonumber A_{ab} & = & \sum_i a_{i} \epsilon_{abcd}\mathcal{L}^c(\alpha_{i})e^d, \\
\nonumber B_{ab} & = & \sum_i b_{i} \epsilon_{abcd}\mathcal{L}^c(\alpha_{i})e^d,  \\
\nonumber C_{ab} & = & C_{ab}^{(1)}+C_{ab}^{(2)}, \\
\nonumber C_{ab}^{(1)} & = & \sum_i c_{i} \epsilon_{abcd}\mathcal{L}^c(\alpha_{i})e^d, \\
\label{cond4} C_{ab}^{(2)} & = & \frac{1}{8 \sigma} \left\{ e_{a} i_{b} \left(-\frac{2\Lambda }{3} d\tau + d\varphi +8\sigma \gamma d\chi \right)  -e_{b} i_{a} \left(-\frac{2\Lambda }{3} d\tau + d\varphi +8\sigma \gamma d\chi \right) \right\}, \,\,\,\,\,\,\,\,\,\,\,
\end{eqnarray}
where $a_i$, $b_i$ and $c_i$, 
and $\alpha_i$, are functions and $i= \tau, \chi, \varphi$.
Then, taking into considerations restrictions (\ref{fe43}), (\ref{fe44}) and (\ref{fe45}) we obtain
\begin{eqnarray}
\nonumber \label{gamma0} 0 & = & -\frac{1}{2\sigma}\left( \gamma d\tau-\frac{\Lambda}{3} d\chi \right) \wedge \epsilon_{abcd} e^a \wedge e^b \wedge e^c \wedge i^d \left( -\frac{2\Lambda}{3} d\tau +d\varphi+8\sigma \gamma d\chi \right)+ d\tau \wedge A_{ab} \wedge e^a \wedge e^b \\
&& +d\chi \wedge B_{ab} \wedge e^a \wedge e^b+d\varphi \wedge C_{ab}^{(1)} \wedge e^a \wedge e^b+ \gamma \epsilon_{abcd} e^a \wedge e^b \wedge e^c \wedge e^d, \,\,\,\,\,\,\,\,\,\, \\
\nonumber 0 & = & \left( \frac{\Lambda^2}{9} + 4 \sigma \gamma^2 \right) \epsilon_{abcd} e^a \wedge e^b \wedge e^c \wedge e^d + 8 \sigma \gamma (d\tau \wedge A_{ab} \wedge e^a \wedge e^b+d\chi \wedge B_{ab} \wedge e^a \wedge e^b \\
\label{gamma1} && +d\varphi \wedge C_{ab}^{(1)} \wedge e^a \wedge e^b)-\frac{2 \Lambda}{3} d\varphi \wedge \epsilon_{abcd} e^a \wedge e^b \wedge C^{(2) cd}, \\
\nonumber 0 & = & -\frac{2 \Lambda}{3} (d\tau \wedge A_{ab} \wedge e^a \wedge e^b+d\chi \wedge B_{ab} \wedge e^a \wedge e^b+d\varphi \wedge C_{ab}^{(1)} \wedge e^a \wedge e^b \\
&& + \gamma \epsilon_{abcd} e^a \wedge e^b \wedge e^c \wedge e^d) +2 \gamma  d\varphi \wedge \epsilon_{abcd} e^a \wedge e^b \wedge C^{(2) cd}
\label{gamma2} ,
\end{eqnarray}
respectively. So, from Eqs. (\ref{gamma0}), (\ref{gamma1}) and (\ref{gamma2}) we devise two possible solutions

\begin{itemize}
\item $\gamma = 0$

when inserting $\gamma=0$ in equations (\ref{gamma0}), (\ref{gamma1}) and (\ref{gamma2}) we obtain $\Lambda=0$ and the constraint (\ref{gamma0}) with $\gamma=0$. Therefore, from Eqs. (\ref{torsiopr}) and (\ref{conto}) the torsion and its associated contortion read
\begin{equation}\label{torsion41}
T^{\left( 4,1 \right)}_{a} =  - \frac{1}{4\sigma }\epsilon _{abcd}e^{b} \wedge e^{c}\wedge i^{d}\left( d\varphi  \right),
\end{equation}
\begin{equation}
{K^{\left( 4,1 \right)}}_{ab}  =  \frac{1}{4 \, \sigma}{\epsilon_{abcd}}\wedge {e^c} \mathcal{L}^d(\varphi).
\end{equation}

\item $\gamma \neq 0$ 

In this case, similarly as in the previous case, we get
\begin{equation}
\nonumber \gamma  = \pm \frac{\Lambda}{6 \sqrt{\sigma}} ,
\end{equation}
which is valid only for the Euclidean case, this is $\sigma=1$. The torsion and contortion read
\begin{eqnarray}
\label{torsion42} T^{\left( 4 , 2\right)}_{a} & = & \frac{\Lambda}{3} \left( d\chi \mp \frac{1}{2} d\tau \right) \wedge e_{a}+ \frac{1}{4 }\epsilon _{abcd}e^{b} \wedge e^{c}\wedge \mathcal{L}^{d}\left(\frac{2\Lambda }{3} \{ \tau \mp  2 \chi \}  - \varphi \right), \,\,\,\,\,\\
\label{contortionfinal42} K^{\left( 4 ,2 \right)}_{ab} & = &  \frac{\Lambda}{3}  \left\{ \mathcal{L}_a \left( \chi \mp \frac{1}{2} \tau \right) e_{b} -\mathcal{L}_b \left( \chi \mp \frac{1}{2} \tau \right) e_{a} \right\}  - \frac{1}{4}\epsilon _{abcd} e^{c}\wedge \mathcal{L}^{d}\left( \frac{2\Lambda }{3} \{ \tau \mp  2 \sigma \chi \}  - \varphi \right), \,\,\,\,\,\,\,\,\,\,\,\,
\end{eqnarray}
respectively. 
\end{itemize}
Note that in solutions (\ref{torsion41}) and (\ref{torsion42}), in the absence of a cosmological constant, the Nieh-Yan coupling $\varphi$ can still generate torsional degrees of freedom by itself.  

\subsection{Matter source} \label{section}

We can add in the action (\ref{action1}) a term representing the matter source:
\begin{equation}
S_{m}= \int \limits_{\mathcal{M}} \mathcal{L}_{m},
\end{equation}
which, when varied with respect to the vierbein, yields the energy-momentum $3$-form $\tau_{a}=\frac{\delta \mathcal{L}_{m}}{\delta e^{a}}=\mathcal{T}_{ab} \wedge \star e^{b}$. Consequently, the term $-2\mathcal{T}_{ab} \wedge \star e^{b}$, where the $\star$ represents the Hodge dual mapping, must be included in the left side of the field equation (\ref{fe41}) or its equivalent in the other cases examined. Additionally, variation with respect to $\omega^{ab}$ yields the spin current $3$-form $\Sigma_{ab}=\frac{\delta \mathcal{L}_{m}}{\delta \omega^{ab}}=\Sigma_{ab}^{\,\,\,\,\, c} \star e_{c}$, thus, the term $-\Sigma_{ab}^{\,\,\,\,\, c} \star e_{c}$ must be included in the left side of the field equation (\ref{fe42}) or its equivalent in the other cases examined.
It is known that the spin current is a source for torsion; however, this is not the only possibility. In the theory considered here, the gradient of the scalar fields $\tau$, $\varphi$ and $\chi$ are also source of torsion as we have shown in the above subsections; so, in our case we can have torsion even if the spin current $\Sigma_{ab}$ is null. In the next section we will consider cosmological scenarios with a standard perfect fluid matter without spin $\Sigma_{ab}=0$.

\subsection{Note}

It is important to note that the particular forms we have considered for the curvature (Eqs. (\ref{cond1}), (\ref{cond2}), (\ref{cond3}) and (\ref{cond4})) may not exhaust all the possible terms that meet the requirements of the curvature $2$-form.

\section{SOME COSMOLOGICAL SOLUTIONS}
\label{cos_sol_sec}


In this section we construct cosmological solutions having in mind the insights gained with the analytical expressions obtained on the previous section. We start then by choosing the following form for the metric
\begin{equation}\label{flrw}
ds^{2}=dt^{2}+\sigma a\left( t\right) ^{2}\left( dx^{2}+dy^{2}+dz^{2}\right),
\end{equation}
where $\sigma =-1$  for a Lorentzian metric, corresponding to the \textit{FLRW} metric with flat spatial curvature, and $\sigma =1$ for an Euclidean metric. We allow for the freedom in signature choice in order to recover interesting solutions that have been reported (for instance the \textit{torsion vortex} metric \cite{Leigh}). The spacetime is assumed to be spatially flat, in accordance with the current observational evidence \cite{Komatsu:2010fb}.  The vierbein is easily obtained to be
\begin{equation}\label{tetrad}
e^{0}=dt,\;\;e^{1}=a\left( t\right) dx,\;\;e^{2}=a\left( t\right)dy,\;\;e^{3}=a\left( t\right) dz,
\end{equation}
notice that this choice is not unique since many other veirbein give rise to the same metric. In fact, any local Lorentz transformation of the diagonal vierbein (\ref{tetrad}) give rise to the \textit{FLRW} metric. However, note that in certain teleparallel theories, such as $f(T)$ gravity, due to their lack of local Lorentz invariance, they become sensitive to the vierbein choice, for a recent review of these ideas see \cite{Cai:2015emx}. From (\ref{originaltorsion}) we obtain the following expressions for the Levi--Civita connection $\bar{\omega}^{ab}$ 
\begin{equation}\label{levicitaexample}
\bar{\omega}^{i0}=He^{i}, \;\;\; \bar{\omega}^{ij}=0, \;\;\; i,j=1,2,3 
\end{equation}
where $H=\frac{a^{\prime }}{a}$ is the Hubble parameter, and $a^{\prime }$ denotes the time derivative of the scale factor $a$.

A cosmological scenario is considered by adding in the action (\ref{action1}) a term representing the matter source (see subsection \ref{section}).
In this work we will consider spinless matter, 
thus $\Sigma_{ab}=0$, and we model the dynamics by a simple pressure-less dust scenario described by a perfect fluid with energy-momentum $2$-form $\mathcal{T}^{a}_{\,\,\,\,b}=(\rho,0,0,0)$, with density $\rho$. Additionally, we argue that the couplings $0$-forms are only dependent on the time coordinate, since, in order to describe realistic solutions, we need to encompass the observational properties of isotropy and homogeneity of our universe at large scales.

Eqs. (\ref{tetrad}) and (\ref{levicitaexample}) are then used to construct the corresponding contortion $K^{\left( i \right)}_{ab}$ and torsion $T^{\left( i \right)}_a$ from the expressions founded in the previous section and the curvature $R^{\left( i \right)}_{ab}$ from the Eq. (\ref{Riemanntensorgral}) with $i = 1,2,3,4$ labeling each of the cases considered in the previous section. We have to recall, though, that the expressions $K_{ab}^{(i)}$ and $T_{a}^{(i)}$ we calculated have been obtained in the absence of matter and thus, must to be modified for the introduction of it. Consequently, having in mind Eq. (\ref{fe41}) with a dust type of matter, it is easy to show that Eq. (\ref{curvature4ansatz}) behaves as follows: $R^{ij}$ acquires the term $\rho /3 e^{i} \wedge e^{j}$ and $R^{0i}$ acquires the term $- \rho /6 e^{0} \wedge e^{i}$ for $i,j=1,2,3$. Also, by using Eq. (\ref{fe42}), one finds that the expression obtained for the torsion is just Eq. (\ref{torsiopr}) with the substitution 
$\Lambda \rightarrow \rho + \Lambda$.

Now, we take a look at the system of differential equations related to the case 4 since it is the one with all the couplings turned on. Using metric (\ref{flrw}), from Eqs. (\ref{torsiopr}) and (\ref{conto}) (with the substitution $\Lambda \rightarrow \rho + \Lambda$) we obtain the expressions for the torsion and contortion, respectively. Then, from (\ref{Riemanntensorgral}) we calculate the curvature $2$-form. Finally, collecting all the field equations (\ref{fe41} - \ref{fe45}) 
we get the following system of equations
\begin{eqnarray}
\label{sistema1} -4\left( H-x\right) +\sigma \tau ^{\prime }xz+4\chi ^{\prime }\left( x^{2}+\frac{\sigma }{16}z^{2}\right) &=& 0,\\
\label{sistema2} z+2\tau ^{\prime }\left( x^{2}+\frac{\sigma }{16}z^{2}\right) -\varphi ^{\prime }+2\chi ^{\prime }xz &=&0, \\
\label{sistema4} 3\left( x^{2}+\frac{\sigma }{16}z^{2}\right) -\Lambda &=&\rho , \\
\label{sistema5} -2\sigma \left( x^{2}+\frac{\sigma }{16}z^{2}\right) -\frac{4\sigma }{a\left( t\right) }\frac{d}{dt}\left( xa\left( t\right) \right) +2\sigma\Lambda +\frac{1}{2}\varphi ^{\prime }z &=&0, \\
\label{sistema6} \frac{3xz}{a\left( t\right)}\frac{d}{dt}\left( za\left( t\right) \right) +\frac{24\sigma}{a\left( t\right)}\left( x^{2}+\frac{\sigma }{16}z^{2}\right) \frac{d}{dt}\left( xa\left( t\right) \right) &=&0, \\
\label{sistema7} -\frac{6xz}{a\left(t\right)}\frac{d}{dt}\left( xa\left( t\right) \right) -\frac{3}{a\left(t\right)} \left( x^{2}+\frac{\sigma }{16}z^{2}\right) \frac{d}{dt}\left( za\left( t\right) \right) &=&0, \\
\label{sistema3} \frac{3}{2}z^{\prime}+\frac{9}{2}Hz &=&0, 
\end{eqnarray}
where we have set 
\begin{eqnarray}
\label{var1} x & = & H-\frac{1}{3}\left( \Lambda +\rho \right) \chi ^{\prime }+\gamma \tau^{\prime }, \\
\label{var2} z & = & -\frac{2}{3}\left( \Lambda +\rho \right) \tau ^{\prime }+\varphi ^{\prime}+8\sigma \gamma \chi^{\prime },
\end{eqnarray}
for convenience. Notice that Eqs. (\ref{sistema6}), (\ref{sistema7}) and (\ref{sistema3}) come from the variation with respect to the Euler, Pontryagin and Nieh-Yan terms respectively, and so are present in the case at hand while not being part of the set of equations when one or more of the characteristic classes are ``switched off'' from the start. 

In fact, all the equations needed to describe the cases we considered in the previous section can be obtained from the system above, recalling that when we  switched off one or more couplings, we did so at the level of the action. This leads to a change in the number of differential equations we then obtain from the action principle, beside of course the vanishing of the corresponding couplings in the remaining equations. We will also cover some cases not presented in section \ref{cas_sec} but that can be obtained from these solutions and can have a cosmological interest.

\subsection{Setting $z=0$ and $\chi=0$} \label{case1}

For this class of solutions we obtain $\gamma \tau ^{\prime}=0$, so either $\gamma=0$ or $\tau=constant$. Having $\tau=constant$ implies $\varphi=constant$, i.e. we recover GR. If we consider $\gamma=0$, the torsion becomes null, and we recover the equations and solutions of GR, too
\begin{eqnarray}
3H^2 & = & \rho+\Lambda ,\\
\rho^{\prime}+3H\rho & = & 0.
\end{eqnarray}

However, the extra structure departing from GR shows itself in the presence of the Pontryagin and Nieh-Yan couplings that satisfy the following equation
\begin{equation} \label{vartau}
2\tau^{\prime}H^2-\varphi^{\prime}=0.
\end{equation}

Note also that if $\tau$ is constant then Eq. (\ref{vartau}) implies that (with $H \neq 0$) $\varphi$ is a constant too, and vice versa. Since the solutions for the scale factor $a \left( t \right)$ for GR are known, we are not considering them here.

\subsection{Setting $\tau =0$ and $\chi =0$} \label{case2}

Although in principle this is not a case that was analyzed in section \ref{cas_sec}, it is clear from the equations obtained that the appearance of just one of the density classes is enough to have torsion. The set (\ref{sistema1} - \ref{sistema7}) thus allows us to study the present configuration which is interesting because of reproducing some models already found in literature. Note that in this case the system (\ref{sistema1} - \ref{sistema7}) looses the equations coming from the couplings $\tau$ and $\chi$ 
leading to
\begin{eqnarray}
\label{sol11} x & = & H ,\\
\label{sol12} z & = & \varphi ^{\prime }  = \frac{A}{a^{3}},
\end{eqnarray}
where $A$ is an integration constant.

Additionally, the non-null components of the torsion tensor are given by
\begin{equation}
T(t) \equiv T_{xyz}=T_{yzx}=T_{zxy}=-T_{xyz}=-T_{yxz}=-T_{zyx}=\frac{\varphi'(t)}{2}a(t)^3,
\end{equation}
and using (\ref{sol12}) we obtain that all of them are constants:
\begin{equation}
T(t)=\frac{A}{2}.
\end{equation}
This torsion corresponds to the $At$ class, according to the classification of the  torsion tensors given in \cite{Capozziello:2001mq} in terms of irreducible tensors in 4 dimensions.

It is worth to notice that considering $\rho \neq 0$ combined with the energy-momentum conservation equation, which is contained in the above system of differential equations, we come to the conclusion that 
\begin{equation}\label{cons}
\rho ^{\prime }+3H\rho =0, \quad \quad \Rightarrow \quad \quad \rho =\frac{\rho _{0}}{a^{3}},
\end{equation}
which in principle represents a cosmological solution. We now present several sub-cases derived from the above

\begin{enumerate}
\item for $\rho =0$, $\sigma =1$ and $\Lambda >0$
\begin{eqnarray}
\label{sol13a} H\left( t\right) & = &\sqrt{\frac{\Lambda }{3}}\tanh \left( \sqrt{3\Lambda }\left( t-t_{0}\right) \right), \\
\label{sol14a} a\left( t\right) & = & a_{0}\left( \cosh \left( \sqrt{3\Lambda }\left( t-t_{0}\right) \right) \right) ^{1/3}, \\
\label{sol15a} \varphi \left( t\right) & = & \frac{2A}{a_{0}^3\sqrt{3\Lambda }} \mathrm{arctan} \left( \mathrm{sech} \left( \frac{\sqrt{3\Lambda }}{2}t\right) \sinh \left( \frac{\sqrt{3\Lambda }}{2}\left( t-2t_{0}\right) \right) \right) ,
\end{eqnarray}
where $t_{0}$ is an integration constant. Note that Eq. (\ref{sol14a}) is exactly the solution known as the torsion vortex \cite{Leigh}. In fact, this solution corresponds to the one given in Eq. (33) of \cite{Leigh}, 
with $e^{A(t)}=a(t)$. Note also that in this paper and for an Euclidean metric ($\sigma=1$), $\Lambda>0$ corresponds to the anti-de Sitter case, because we have written the cosmological constant term in the action (\ref{EHaction}) with a plus sign. This choice was done in order that for a Lorentzian metric ($\sigma =-1$) with signature $(+,-,-,-)$ $\Lambda<0$ corresponds to the anti-de Sitter 
case.

\item for $\rho \neq 0$, $\sigma =-1$ and $\Lambda < 0$, we obtain the following solution


\begin{eqnarray}
\nonumber a\left( t\right) & = & \frac{1}{2^{4/3}\left( -\Lambda \right) ^{1/2}}e^{\sqrt{-\frac{\Lambda}{3}}\left( t-t_{0}\right)}\left(-12 \Lambda A^2-\left(e^{-\sqrt{-3\Lambda }\left( t-t_{0}\right) }\sqrt{-\Lambda } +12B \right)^2\right) ^{1/3}, \\
\label{aa}\\
\label{ab} \varphi \left( t\right) & = & -\frac{8}{3} \mathrm{arccoth} \left( \frac{2A\Lambda\sqrt{3}}{ \Lambda e^{-\sqrt{-3\Lambda }\left( t-t_{0}\right) }-12B\sqrt{-\Lambda }}\right), \\
\label{ac} \rho \left( t\right) & = & -\frac{3B}{a^3\left(t\right)},
\end{eqnarray}
with $t_{0}$ and $B<0$ integration constants. In Fig. \ref{Figure} we show this set of solutions of $a(t)$, $T(t)$, $q(t)$ and 
$\varphi(t)$ for $A=10$, $B=-1$, $t_0=0$ and $\Lambda=-1$, where $q(t)$ denotes the deceleration parameter, defined by
\begin{equation}
q(t)=-\frac{a(t) a''(t)}{(a'(t))^2}.
\end{equation} 

It is interesting to note that the coupling zero-form $\varphi \left( t \right)$ is asymptotically slowly varying after $t \approx 1$. Consequently, the influence of torsion $T_a$ over the field equations becomes increasingly negligible for $ t > 1 $. Note also how the scale factor $a \left( t \right)$ is non-decreasing, thus, suggesting a non-contracting universe. This is even clearer when examining Fig. \ref{Figure1}, where we show the behavior of $q$, $\varphi$ and $T$, all of them as a function of $a(t)$. We observe that the deceleration parameter is initially positive, then becomes negative and tends to a constant value, describing an accelerated expanding  universe at later times.
\begin{figure}[!h]
\centering
\includegraphics[width=4in,height=3in]{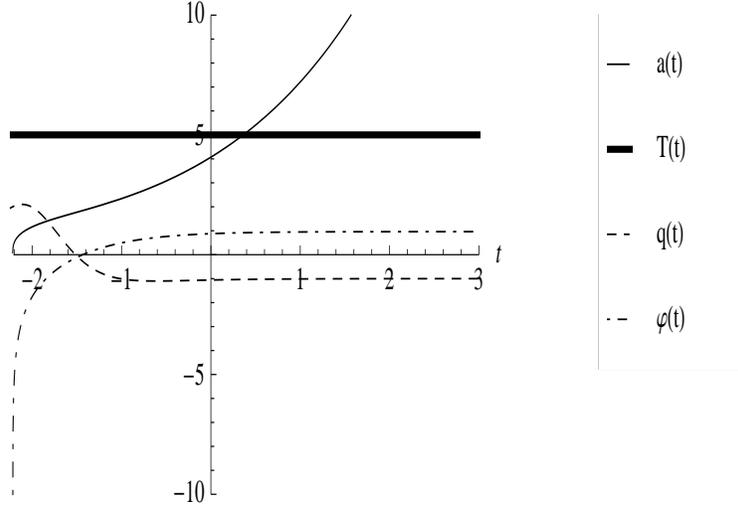}
\caption{\label{Figure} Behavior of $a(t)$, $T(t)$, $q(t)$ and $\varphi(t)$ for $A=10$, $B=-1$, $t_0=0$ and $\Lambda=-1$}
\end{figure}

\begin{figure}[!h]
\centering
\includegraphics[width=4in,height=3in]{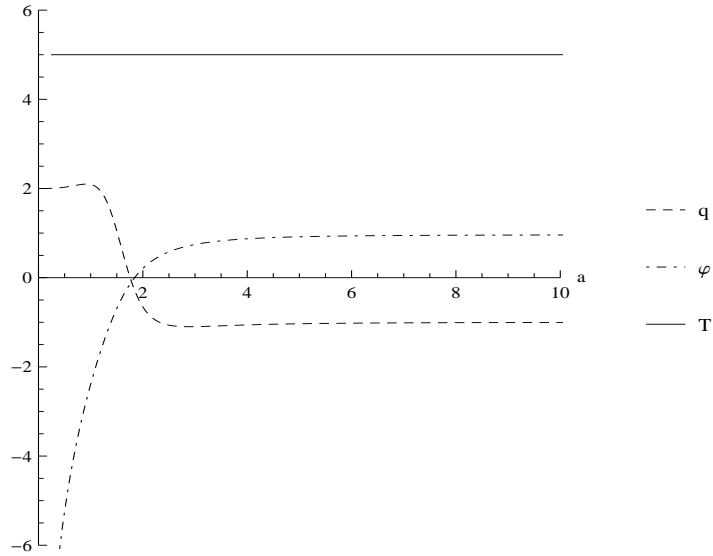}
\caption{\label{Figure1} Behavior of $q$, $\varphi$ and $T$ as function of $a$, for $A=10$, $B=-1$, $t_0=0$ and $\Lambda=-1$}
\end{figure}

\end{enumerate}

\begin{enumerate}
\item for $\rho = 0$, $\sigma =-1$ and $\Lambda > 0$, the cosmological solutions coincide with those of equations (\ref{sol13a} - \ref{sol15a}), where we have plotted in Fig. \ref{Figure2} the set of solutions for $a(t)$, $T(t)$, $q(t)$ and $\varphi(t)$ when $a_0=1$, $A=1$, $t_0=0$, $\Lambda=1$. We can see again how the coupling zero-form $\varphi$ is asymptotically slowly varying after $t \approx 1$, hence, with negligible contribution in the field equations. This is better shown in Fig. \ref{Figure3} where the behavior of $q$, $\varphi$ and $T$ as a function of $a(t)$ has been plotted. Notice the two branched, asymptotically slow varying solution for $\varphi$, as well as having a deceleration parameter that is initially negative, but that later increases monotonically to a constant negative value.

\begin{figure}[!h]
\centering
\includegraphics[width=4in,height=3in]{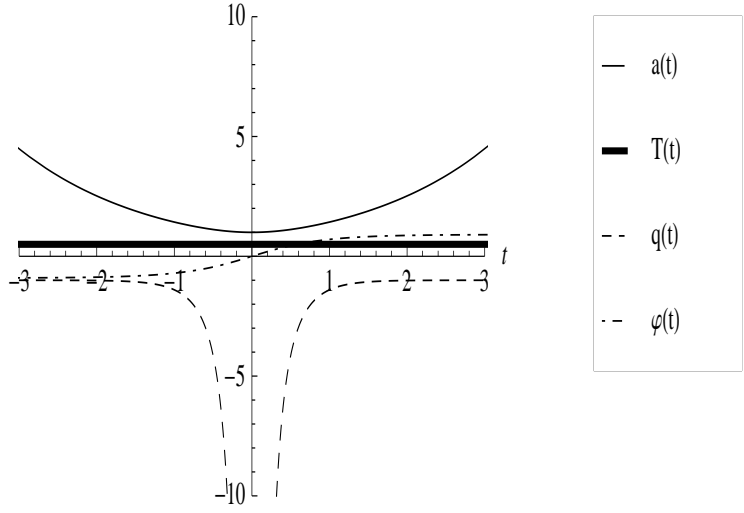}
\caption{\label{Figure2} Behavior of $a(t)$, $T(t)$, $q(t)$ and $\varphi(t)$ for $a_0=1$, $A=1$, $t_0=0$ and $\Lambda=1$}
\end{figure}

\begin{figure}[!h]
\centering
\includegraphics[width=4in,height=3in]{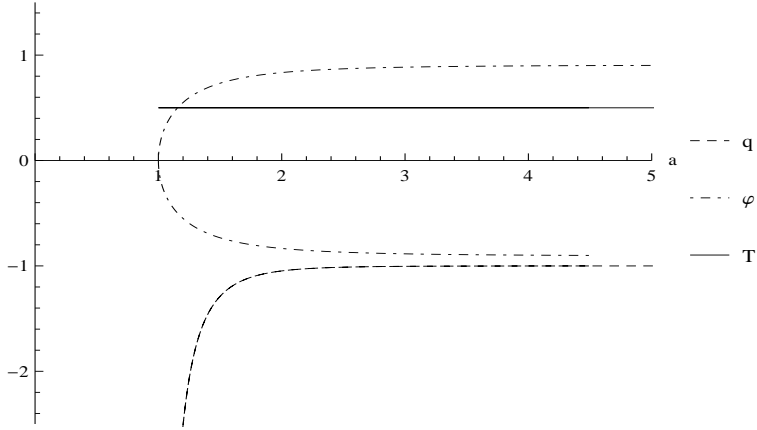}
\caption{\label{Figure3} Behavior of $q$, $\varphi$ and $T$ as function of $a$, for $a_0=1$, $A=1$, $t_0=0$ and $\Lambda=1$}
\end{figure}

\item for $\rho = 0$, $\sigma =-1$ and $\Lambda < 0$ we have
\begin{eqnarray}
\label{sol13d} a \left( t\right) & = & a_{0} \left( \cos \left( \sqrt{-3\Lambda }\left( t-t_{0}\right)\right) \right) ^{1/3}, \\
\nonumber \varphi \left( t\right) & = & \frac{2A}{\sqrt{-3\Lambda }a_{0}^3} \mathrm{arctanh} \left( \sec \left( \frac{\sqrt{-3\Lambda }}{2}t\right) \sin \left( \frac{\sqrt{-3\Lambda }}{2}\left( t-2t_{0}\right) \right) \right).\\
\label{sol14d}
\end{eqnarray}
In order to be physically admissible, at least as a model of the early stages of cosmic evolution, we must restrict the cosmic time to the interval $-\frac{\pi}{2\sqrt{-3\Lambda}}\leq t \leq 0$, with $t_0=0$. In this way $a(t)$ is an increasing function of time. From equation (\ref{sol13d}), it is easily seen that the deceleration parameter yields
\begin{eqnarray}
\label{sol15d} q \left( t \right) & = & \frac{\Lambda}{3} \left\{1 + \sec^{4/3} \left( \sqrt{-3 \Lambda } \left( t - t_o \right) \right) \csc^2 \left( \sqrt{-3 \Lambda } \left( t - t_o \right) \right) \right\}. \quad 
\end{eqnarray}

\item for $\rho = 0$, $\sigma =-1$ and $\Lambda = 0$ we obtain
\begin{eqnarray}
\label{sol13e} a \left( t\right) & = & a_{0} t^{1/3}, \\
\label{sol14e} \varphi \left( t\right) & = & \frac{A}{a_{0}^3}\log \left( t\right).
\end{eqnarray}
\end{enumerate}
This solution illustrates two monotonically increasing functions for $a \left( t \right)$ and $\varphi \left( t \right)$. It is straightforward from equation (\ref{sol13e}), it easily follows that
\begin{eqnarray}
\label{sol15e} q \left( t \right) & = & 2,
\end{eqnarray}
and hence, represents a decelerated expanding universe.

\subsection{Setting $\chi =0$ and $ \varphi =0$} \label{case3}

As before, since only one coupling is needed to obtain torsion, we go to analyze the case in which only the Pontryagin class is present. When discarding equations coming from the Nieh-Yan invariant constraint and from the Euler invariant constraint, it can be shown that in the remaining system (\ref{sistema1} - \ref{sistema7}) not all equations are independent. Additionally, the non-null components of the torsion tensor are given by
\begin{eqnarray}
\nonumber &&T_1(t)\equiv T_{xtx}=T_{yty}=T_{ztz}=-T_{xxt}=-T_{yyt}=-T_{zzt}=\gamma \tau'(t) a(t)^2,\\
\nonumber &&T_2(t)\equiv T_{xyz}=T_{yzx}=T_{zxy}=-T_{xzy}=-T_{yxz}=-T_{zyx}=-\frac{\Lambda}{3} \tau'(t)a(t)^3, \\
\end{eqnarray}
tuning these into the $Vt$ and $At$ classes of torsion, according to the classification given in \cite{Capozziello:2001mq}.

Several solutions are presented in what follows. First, for $\tau^{\prime}=0$ we recover the solutions of GR:

\begin{enumerate}
\item for $\rho = 0$ and $\Lambda >0$
\begin{eqnarray}
H\left( t\right) & = & \sqrt{\frac{\Lambda}{3}}, \\
a\left( t\right) & = & a_{0}e^{\sqrt{\frac{\Lambda}{3}}t}.
\end{eqnarray}

Representing a dark-energy dominated universe in standard cosmology.

\item for $\rho \neq 0$

\begin{itemize}

\item $\Lambda<0$

\begin{eqnarray}
\label{sol13.a} H\left( t\right) & = &\sqrt{-\frac{\Lambda }{3}}\tan \left( -\frac{\sqrt{-3\Lambda }}{2}\left( t-t_{0}\right) \right), \\
\label{sol14.a} a\left( t\right) & = & a_{0}\left( \cos \left( \frac{\sqrt{-3\Lambda }}{2}\left( t-t_{0}\right) \right) \right) ^{2/3}, 
\end{eqnarray}
yielding a deceleration parameter
\begin{eqnarray}
\label{sol15.a} q \left( t \right) & = & \frac{\Lambda}{12} \csc^2 \left(\frac{\sqrt{-3 \Lambda}}{2} \left( t - t_0 \right) \right).
\end{eqnarray}

\item $\Lambda>0$

\begin{eqnarray}
\label{sol13.b} H\left( t\right) & = &\sqrt{\frac{\Lambda }{3}}\tanh \left( \frac{\sqrt{3\Lambda }}{2}\left( t-t_{0}\right) \right), \\
\label{sol14.b} a\left( t\right) & = & a_{0}\left( \cosh \left( \frac{\sqrt{3\Lambda }}{2}\left( t-t_{0}\right) \right) \right) ^{2/3}. 
\end{eqnarray}
yielding a deceleration parameter
\begin{eqnarray}
\nonumber  q \left( t \right) & = & -\frac{9 \Lambda}{8} \tanh^2 \left( \sqrt{3 \Lambda} \left(t-t_0 \right) \right) \{\frac{1}{3} \textrm{sech}^2 \left( \sqrt{3 \Lambda} \left(t-t_0 \right) \right)  +\\
\label{sol15.b} & & +\textrm{sech}^{\frac{4}{3}} \left( \sqrt{3 \Lambda} \left(t-t_0 \right) \right) +\frac{1}{3} \}.
\end{eqnarray}

\item $\Lambda=0$
\begin{eqnarray}
\label{sol13.c} H\left( t\right) & = &\frac{2}{3t-t_{0}}, \\
\label{sol14.c} a\left( t\right) & = &\left( 3t-t_{0}\right) ^{2/3},
\end{eqnarray}
which it is known to correspond to a matter dominated universe in standard cosmology.
\end{itemize}
\end{enumerate}

For $\tau^{\prime} \neq 0$ and $\rho=0$, from the system of equations one can obtain the following equations
\begin{equation} \label{H}
H=-x+\frac{6}{\Lambda }x^{3},
\end{equation}
\begin{equation}
x^{\prime }-x^{2}+\frac{6}{\Lambda }x^{4}-\frac{\Lambda }{3}=0,
\end{equation}

and from this last equation we can only obtain an implicit solution for $x(t)$:

\begin{enumerate}

\item $\Lambda >0$
\begin{equation}
\frac{1}{\sqrt{3\Lambda }}\left( \mathrm{arctanh} \left( \sqrt{\frac{3}{\Lambda }} x \right) +\sqrt{2}\arctan \left( \sqrt{\frac{6}{\Lambda }}x\right) \right) = t - t_{0}.
\end{equation}

\item $\Lambda <0$
\begin{equation}
-\frac{1}{\sqrt{-3\Lambda }}\left( \arctan \left( \sqrt{-\frac{3}{\Lambda }} x \right) +\sqrt{2}\mathrm{arctanh} \left( \sqrt{-\frac{6}{\Lambda }}x\right) \right) = t - t_{0}.
\end{equation}

\item $\Lambda =0$
\begin{equation}
a=a_{0}.
\end{equation}

\end{enumerate}

It would be worth to study numerically the solutions obtained in this case and also study the case $\tau^{\prime}\neq 0$, $\rho \neq 0$.

\subsection{Setting $\varphi =0$ and $\tau =0$} \label{case4}

As before, this is a case that can be studied starting from the system (\ref{sistema1} - \ref{sistema7}). This was not a case considered in section \ref{cas_sec}, however, it shows some interesting features. When discarding those equations coming from Nieh-Yan invariant constraint and from the Pontryagin invariant constraint, the remaining system contains the conservation Eq. (\ref{cons}). In this case, the non-null components of the torsion tensor are given by
\begin{eqnarray}
\nonumber &&T_1(t)\equiv T_{xtx}=T_{yty}=T_{ztz}=-T_{xxt}=-T_{yyt}=-T_{zzt}=-\frac{\Lambda}{3} \chi'(t) a(t)^2,\\
\nonumber &&T_2(t)\equiv T_{xyz}=T_{yzx}=T_{zxy}=-T_{xzy}=-T_{yxz}=-T_{zyx}= -4 \gamma \chi'(t)a(t)^3,\\
\end{eqnarray}
that can be classified as $Vt$ and $At$ classes of torsion, according to \cite{Capozziello:2001mq}.

After solving the remaining system the following admissible solutions are derived

\begin{enumerate}


\item for $\rho =0$ and $\Lambda >0$
\begin{eqnarray}
\label{sol42a} x\left( t\right) & = &\sqrt{\frac{\Lambda }{2}}\tanh \left( \sqrt{\frac{\Lambda }{2}}\left( t-t_{0}\right) \right), \\
\label{sol42b} \chi \left( t\right) & = & -\frac{1}{\Lambda }\log \left( \sinh \left( \sqrt{\frac{\Lambda }{2}}\left( t-t_{0}\right) \right) \right), \\
a\left( t\right) & =& a_{0}\frac{\cosh \left( \sqrt{\frac{\Lambda }{2}}\left( t - t_{0}\right) \right) }{\left( \sinh \left( \sqrt{\frac{\Lambda }{2}} \left( t-t_{0}\right) \right) \right) ^{1/3}}.
\end{eqnarray}

In Fig. \ref{Figure4} we show this set of solutions of $a(t)$, $T_1(t)$, $T_2(t)$, $q(t)$ and 
$\chi(t)$ for $a_0=1$, $t_0=0$ and $\Lambda=1$. Note that $T_2$ becomes imaginary for $t < 1.63$, this makes the solution valid only for $t > 1.63$.We see that the scale factor increases monotonically after $t \approx 1.5$, suggesting an expanding universe at later times. We also see that the coupling zero-form $\chi$ is decreasing monotonically. This can be further seen in Fig. \ref{Figure5}, where we show the behavior of $q$, $\chi$ and $T$ as a function of $a(t)$. Note the two branching solution of the coupling zero-form $\chi$ and how the deceleration parameter stabilizes in a constant negative number, meaning that the universe described by this model is an expanding universe.

\begin{figure}[!h]
\centering
\includegraphics[width=4in,height=3in]{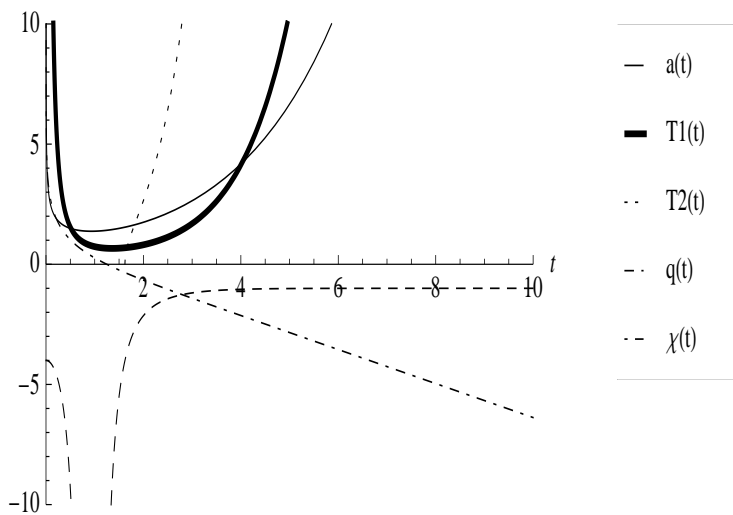}
\caption{\label{Figure4} Behavior of $a(t)$, $T_1(t)$, $T_2(t)$, $q(t)$ and $\varphi(t)$ for $a_0=1$, $t_0=0$ and $\Lambda=1$}
\end{figure}

\begin{figure}[!h]
\centering
\includegraphics[width=4in,height=3in]{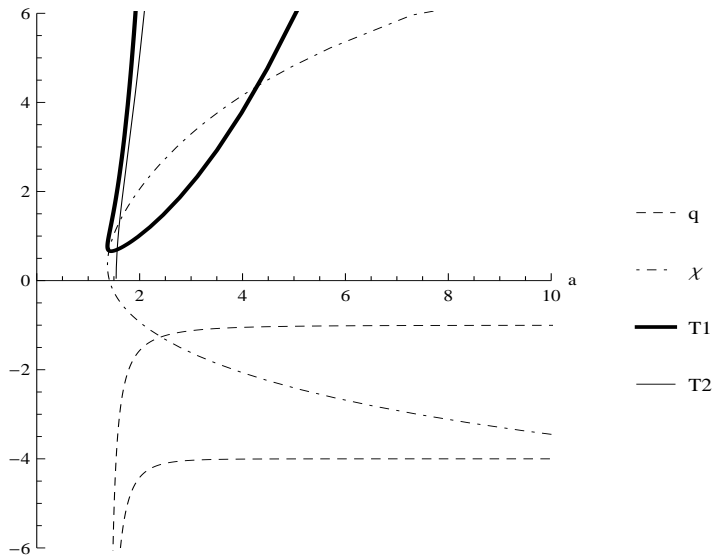}
\caption{\label{Figure5} Behavior of $q$, $\chi$, $T_1$ and $T_2$ as function of $a$, for $a_0=1$, $t_0=0$ and $\Lambda=1$}
\end{figure}

\item for $\rho =0$ and $\Lambda =0$
\begin{eqnarray}
\label{sol43a} x\left( t\right) & = & \frac{1}{ t - t_{0}}, \\
\label{sol43b} \chi \left( t\right) & = & -\frac{t\left( t-2t_{0}\right)}{4}, \\
a\left( t\right) & = & t-t_{0}.
\end{eqnarray}

In this case $a(t)$ increases linearly with time. Notice that an Euclidean metric with $\sigma =1$ only satisfies the condition $\Lambda > 0$.
\end{enumerate}

We do not show the behavior of the solutions for a dust type of matter since they were studied in \cite{Toloza:2012sn, Toloza}.

\section{FINAL REMARKS}

In this work, we studied gravitational theories with propagating torsion, which arises by adding to the Einstein-Hilbert action combinations of the characteristic densities of a space-time manifold coupled with scalar fields. We have found explicit expressions for the contortion $1$-form, given by Eqs. (\ref{contortionfinal1}), (\ref{contortionfinal2}), (\ref{contoprevia}) and (\ref{conto}), consequently, expressions for the torsion (\ref{torsion1}), (\ref{torsion2}), (\ref{torsion3previa}) and (\ref{torsiopr}) for a broad class of situations. These expressions were obtained in terms of the vierbein, the scalar fields (couplings) and their derivatives, only. One interesting feature of such analytical expressions is that important terms depend linearly on the cosmological constant, hence, since observation suggests the existence and positivity of this term, it is tempting to theorize that torsional degrees of freedom could be eventually measured. On the other hand, the strict dependency on the scalar fields shows that torsion becomes null when the couplings are null or constants, recovering the solution of Einstein-Cartan gravity as expected.

Since the cases of section \ref{cas_sec} were studied in a coordinate free fashion, another alternative to encompass the task of study the material presented here is to consider a different metric than the one used here. This program will allow to find some known solutions as well as possibly new ones and will be considered in future works.

On the other hand, by taking a metric such as the one showed in (\ref{flrw}), compatible with the topological assumptions that a space-time is a closed manifold, we were able to reproduce some well known models in a more or less general framework, for instance the torsion vortex solution reported in \cite{Leigh}. Furthermore, we obtained analytically new classes of cosmological solutions for the analyzed cases. These solutions describe physically relevant space-times representing different stages of cosmic evolution. Much more work is certainly necessary to fully study cosmological solutions and it will be tackled in subsequent papers, as well as considering the presence of different kinds of matter fields. Probably, by doing so, the analyticity of the solutions will be lost for which the use of numerical techniques is to be expected.


\section{Acknowledgements}

The authors want to thank Gast\'{o}n Giribet for his comments and valuable suggestions for starting this work. The authors also want to thank Michele Fontanini for his valuable corrections, comments and suggestions in the final steps of this work. This work was funded by the Comisi\'{o}n Nacional de Ciencias y Tecnolog\'{i}a through FONDECYT Grant 11121148 (Y.V., J.L.). J.L. acknowledge the hospitality of the Universidad de La Serena where part of this work was carried out.












\end{document}